\documentclass{article}

\usepackage{PRIMEarxiv}
\usepackage[utf8]{inputenc}
\usepackage[T1]{fontenc}
\usepackage{hyperref}
\usepackage{url}
\usepackage{booktabs}
\usepackage{amsfonts}
\usepackage{nicefrac}
\usepackage{microtype}
\usepackage{lipsum}
\usepackage{fancyhdr}
\usepackage{graphicx}
\usepackage{listings}
\graphicspath{{media/}}
\usepackage{caption}
\usepackage{subcaption}
\usepackage{amsmath}
\usepackage{float}

\title{
MATLAB Plasmonic Nanoparticle Virion Counting and Interpretation System in Urban Populations

}

\author{
Bryan Hong \\
  Independent Researcher \\
  \texttt{bryan.hong.research@gmail.com}
  \and
  Jai Pal \\
  Independent Researcher \\
  \texttt{jaipal9621@gmail.com}
}

\begin{document}
\maketitle

\begin{abstract}
One of the biggest issues currently plaguing the field of medicine is the lack of an accurate and
efficient form of disease diagnosis especially in urban settings such as major cities. For example, the two most commonly utilized test diagnosis systems, the PCR and rapid test, sacrifice either accuracy or speed to achieve the other, and this could slow down epidemiologists working to combat the spread. Another issue currently present is the issue of viral quantification or the counting of virions within a nasal sample. These can provide doctors with crucial information in treating infections; however, the current mediums are underdeveloped and unstandardized. This project’s goals were to 1) create an accurate and rapid RSV diagnostic test that could be replicated and utilized efficiently in urban settings and 2) design a viral quantification mechanism that counts the number of virions to provide more information to healthcare workers. This diagnostic test involved a system that pumped RSV-aggregated Au-nanoparticles and unaggregated Au-nanoparticles through a microcapillary, whose cross-section was intersected by two laser beams generating and detecting the nanobubbles. The signals between the unaggregated and aggregated nanobubbles were calibrated, and the number of RSV virions was recorded. The results yielded an accuracy of 99.99\% and an average time of 5.2 minutes, validating that this design is both faster and more accurate compared to current tests. When cross-validated with Poisson statistics, the virion counting system counted the number of virions with 98.52\% accuracy. To verify the accuracy of our samples, the results were compared to clinical trials of nasal samples, and our diagnostic system predicted accurate
diagnostics after statistical analysis. With further testing, this diagnostic method could replace current standards of testing, saving millions of lives every year. 

\end{abstract}

\section{Introduction}
After crippling society for the last two years, the COVID-19 pandemic was an important wake-up call for society, exposing our need for a more effective system in promising protection for its citizens against viruses. One major cause of this burden is the flaws in our current diagnostic methods. Both nucleic acid tests and antibody tests have major issues and technologically lag behind our current medical advancements. For example, the nucleic acid test can deliver accurate results with a \(97\%\) success rate; however, they take from \(24-72\) hours for results to be accurately evaluated. The antibody test, on the other hand, delivers rapid results in just over \(15\) minutes; however, they only present a \(75\%\) success rate in correctly diagnosing the patient, according to a study conducted by the University of Massachusetts.

Furthermore, our world has realized the urgent need for a more advanced and sophisticated system of diagnosis that allows for more information on the virus sample to be collected during diagnosis. One example of this is viral quantification. Viral load quantification provides scientists with important analytics and insights about the viral load that unlock our ability to understand things such as the prognosis of a virus and the severity of symptoms. For example, by analyzing cycle thresholds in COVID-19 test results, scientists are able to connect and gather data about infection trends and other critical pieces of information to better personalize and design treatments for specific patients.

However, our current means of viral quantification involving cycle thresholds create too much variability and aren't standardized enough to be utilized effectively. Today’s research experiment uses the Respiratory Syncytial Virus (RSV) as a model target for the virus diagnosis in this experiment, but the system designed today can be calibrated and replicated for other viruses of this sort. Historically, the Human Respiratory Syncytial Virus (RSV) has been attributed as the major respiratory pathogen in young children and infants.

\begin{figure}[H]
    \centering
    \includegraphics[width=0.5\linewidth]{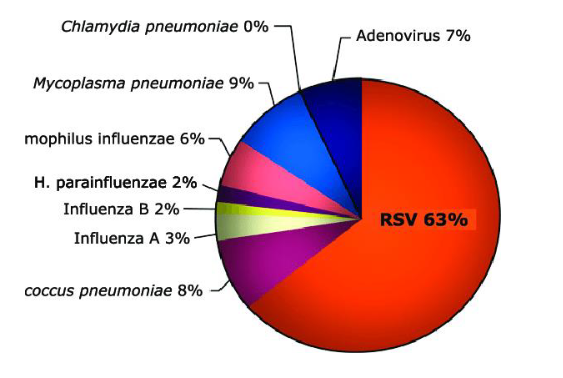}
    \caption{Etiology of acute respiratory infections in children worldwide}
    \label{fig:1}
\end{figure}

\section{Urban Settings}

Urban settings, with their dense populations and intricate social dynamics, present unique challenges in the realm of public health. The management and prevention of infectious diseases in such environments are particularly critical, given the rapidity and scale at which pathogens can spread. Recent data highlights a concerning trend: urban areas, owing to their high population densities, tend to experience earlier and more pronounced peaks in Respiratory Syncytial Virus (RSV) rates compared to less populated regions. This paper introduces an innovative project aimed at addressing healthcare difficulties in urban settings, particularly focusing on mitigating the impact of RSV.

\begin{figure}[H]
    \centering
    \includegraphics[width=0.5\linewidth]{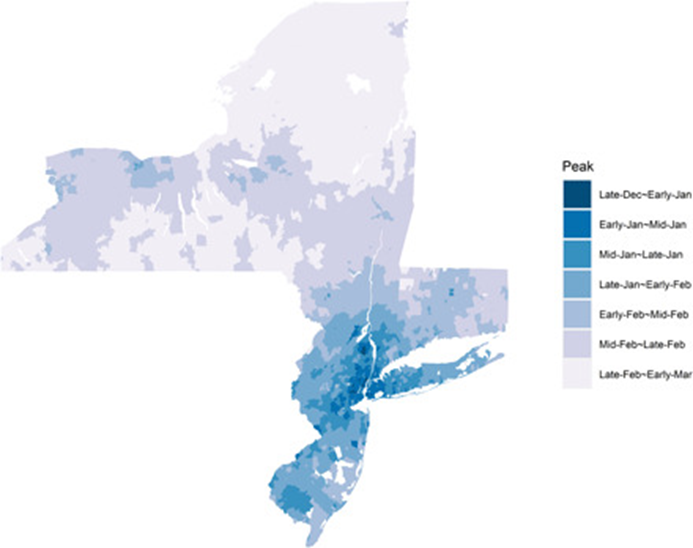}
    \caption{Progression of RSV in rural and urban areas}
    \label{fig:1}
\end{figure}

The figure shows that areas with higher populations, urban cities, typically have earlier and higher peaks for their RSV rates. The aim of this research project is to develop a system that will be able to identify the spread of a virus faster, more efficiently, more accurately, and accessible in urban cities.

\section{Engineering Goal}
The primary objective of this project is to create a sensitive Human Respiratory Syncytial Virus (RSV) diagnostic algorithm using a digital plasmonic nanobubble photodetection system that can be replicated and calibrated for diagnosing other viruses such as, but not limited to: SARS-COV-2, Influenza A \& B, different strains of the Rhinovirus, Human Parainfluenza Viruses, and other types of respiratory viruses.

Unlike our current gold standard, the PCR test, this diagnostic method retrieves rapid and accurate results without the need for a virus amplification step, allowing us to save resources and time. Compared to our current technology of antigen rapid test kits, this diagnostic method retrieves results both more rapidly and accurately.

Additionally, this project utilizes a MATLAB coding algorithm to conduct a single virion counting system that allows scientists to identify the severity of the symptoms and predict the prognosis of the RSV sample.

\section{Materials}

The materials utilized to create this project were divided into three categories: chemical, biological, and mechanical. The chemical materials include \(98 \, \text{mL}\) of Deionized Water (\(dH_2O\)) with a resistivity of \(18.2 \, \text{M}\Omega \cdot \text{cm}\), \(0.338 \, \text{mL}\) of \(2.23 \, \text{nm}\) AuNP seeds, \(1 \, \text{mL}\) of \(99\%\) \(25 \, \text{mM}\) Tetrachloroauric (III) acid trihydrate (\(HAuCl_4 \cdot 3H_2O\)), \(1 \, \text{mL}\) of \(99\%\) \(112.2 \, \text{mM}\) Sodium Citrate tribasic dihydrate (\(Na_3CA \cdot 2H_2O\)), Sodium Chloride (\(NaCl\)), \(50 \, \text{mg}\) of \(5 \, \text{mM}\) 3,3'-Dithiobis(sulfosuccinimidyl propionate) (DTSSP), and a \(2 \, \text{mM}\) Borate buffer bath.

The biological materials include samples of Synagis (Palivizumab) and strains of purified A2 Human Respiratory Syncytial Virus (RSV).

The mechanical tools used in this experiment include a \(24 \, \text{slot}\) centrifuge machine, \(250 \, \text{mL}\) Erlenmeyer flasks, a magnetic hot plate, AmiconTM ultra centrifugal filter units, a \(28 \, \text{picosecond}\) \(532 \, \text{nm}\) PL 2230 Ekspla pulse laser, a \(633 \, \text{nm}\) Newport red HeNe continuous laser, a photodetector to detect changes in \(mV\) from the laser beam, a light filter wheel to adjust the intensity of the laser beam, mirrors and apertures to focus and align the beams, a \(200 \, \mu \text{m}\) microcapillary, a Synergy 2 BioTek plate reader, a Malvern ZetaSizer Nano ZS DLS Machine, a \(10 \, \text{K MWCO}\) Dialysis Cassette, a JEOL JEM-2010 transmission electron microscope, a syringe pump, a variety of different-sized pipets, an oscilloscope for data collection, and a computer with the MATLAB software for data collection.

\section{Synthesis of the 15 nm AuNPs}
The Plech Turkevich method was utilized to synthesize the \(15 \, \text{nm}\) AuNPs. The \(15 \, \text{nm}\) AuNPs were used to conjugate with the Synagis (Palivizumab) antibodies to attach and conjugate the RSV virion samples.

First, \(98 \, \text{mL}\) of \(dH_2O\) and \(1 \, \text{mL}\) of the \(25 \, \text{mM}\) Tetrachloroauric (III) acid trihydrate were mixed inside a pre-cleaned \(250 \, \text{mL}\) Erlenmeyer flask on a magnetic hot plate. The solution was mixed vigorously and heated to a boil. Next, \(1 \, \text{mL}\) of the \(112.2 \, \text{mM}\) Sodium Citrate tribasic dihydrate was injected into the flask using a \(1000 \, \mu \text{L}\) pipet. Finally, the flask was removed from the magnetic hot plate after a color change occurred in the flask, indicating that the reaction has been completed.

After the process was completed, spectral absorbance of the \(15 \, \text{nm}\) AuNP samples were measured by a Synergy 2 BioTek plate reader, and their hydrodynamic size was measured and checked using a Malvern ZetaSizer Nano ZS DLS machine. This step was then repeated multiple times to create multiple batches for the different trials.

\section{Conjugation of Antibodies and RSV with the AuNPs}
The Synagis (Palivizumab) antibodies were conjugated with one batch of the \(15 \, \text{nm}\) AuNPs. The antigen binding sites of the Synagis (Palivizumab) antibodies target and bind to the RSV surface F glycoproteins, allowing the RSV virions to be conjugated with the AuNPs. The DTSSP crosslinker was linked to the surface of the AuNP and links the Synagis (Palivizumab) antibodies to the AuNPs. During this step of the process, the amine-reactive N-hydroxysulfosuccinimide (sulfo-NHS) ester group at the two ends of the DTSSP reacts with the amine groups on the Synagis (Palivizumab) antibodies at pH 7-9, forming stable amide bonds. The DTSSP contains a disulfide bond in the center of the compound (Figure 3). Through hydrolysis, the disulfide bond is able to be split apart into two separate identical parts. After the sulfo-NHS-ester bonds react with the primary amine bonds in antibodies and bond together, the single sulfur bonds at the center of the DTSSP crosslinking reagent will form covalent bonds with the sulfurs on the surface of the AuNP.
\begin{figure}[H]
    \centering
    \includegraphics[width=0.5\linewidth]{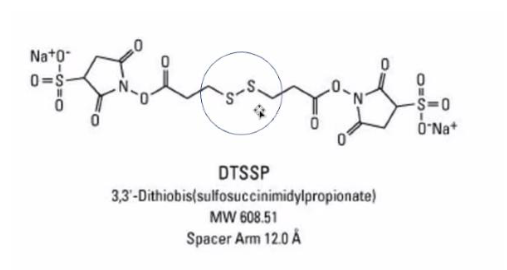}
    \caption{Chemical Structure of a DTSSP crosslinking reagent}
    \label{fig:2}
\end{figure}

First, \(1 \, \text{mL}\) of \(15 \, \text{nm}\) AuNPs were washed by a high-speed centrifugal machine with \(10,000 \, g\) for \(25 \, \text{minutes}\) before being added back to the \(2 \, \text{mM}\) borate buffer (\(pH = 8.5\)). Next, \(5 \, \text{mM}\) of the DTSSP was added to the Synagis (Palivizumab) with a molar ratio of \(125:1\). Then, the DTSSP-Synagis solution was injected into a \(10 \, \text{K MWCO}\) Dialysis Cassette and dialyzed for \(4 \, \text{hours}\) before being transported into \(100 \, \text{kDa AmiconTM}\) centrifugal filters to remove and clean out the unlinked DTSSP. The resulting DTSSP-Synagis solution was then added to the \(15 \, \text{nm}\) AuNPs in the \(2 \, \text{mM}\) borate buffer. Finally, the solution was kept in an ice bath for \(2 \, \text{hours}\) before getting washed by the centrifuge machine for a couple of times. The \(15 \, \text{nm}\) AuNPs were now conjugated and linked together with the DTSSP-Synagis link, forming an AuNP-DTSSP-Synagis link. The AuNP-DTSSP-Synagis was then stored in the \(2 \, \text{mM}\) borate buffer at \(4 \, ^\circ \text{C}\).

This process was repeated for two steps because one batch was utilized as the calibration control to determine the threshold for the single-AuNP counting mechanism, while the other batch was further conjugated with the purified RSV virus and utilized in the single-virion counting mechanism. One batch of the AuNP-DTSSP-Synagis solution was cleaned and prepared to be conjugated with the purified RSV strains. The purified RSV strains were incubated with the AuNP-DTSSP-Synagis in the \(2 \, \text{mM}\) Borate buffer at room temperature. After \(30 \, \text{minutes}\), the solution was removed.

\begin{figure}[H]
    \centering
    \includegraphics[width=0.5\linewidth]{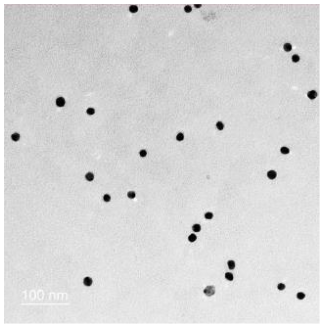}
    \caption{15 nm AuNP-DTSSP-Synagis probes}
    \label{fig:3}
\end{figure}
\begin{figure}[H]
    \centering
    \includegraphics[width=0.5\linewidth]{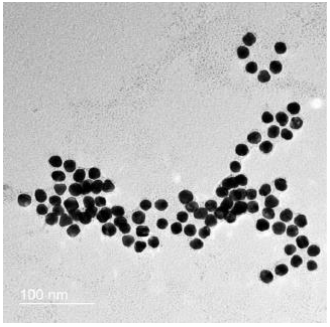}
    \caption{RSV-conjugated 15 nm AuNP-DTSSP-Synagis probes}
    \label{fig:4}
\end{figure}

\section{Detection System Setup}
The setup for the plasmonic nanobubble detection system consists of a \(532 \, \text{nm}\) \(28\)-picosecond pulse laser and a \(633 \, \text{nm}\) red Helium Neon continuous laser that is aligned through the cross-section of a \(200 \, \mu \text{m}\) microcapillary (Figure 5) and then focused into a photodetector using apertures and mirrors. The energy from the laser beams is collected using the photodetector and an oscilloscope (Figure 6).

\begin{figure}[H]
    \centering
    \includegraphics[width=0.5\linewidth]{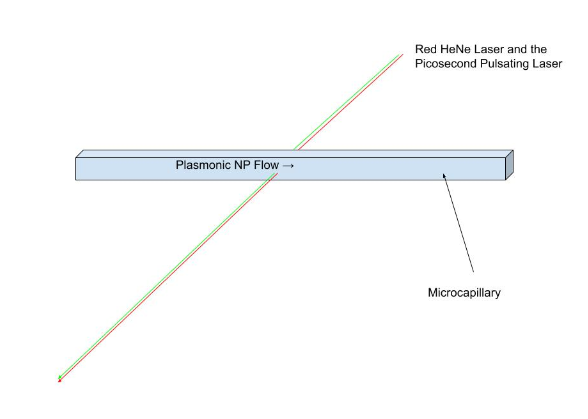}
    \caption{Microcapillary alignment with the lasers}
    \label{fig:5}
\end{figure}

\begin{figure}[H]
    \centering
    \includegraphics[width=0.5\linewidth]{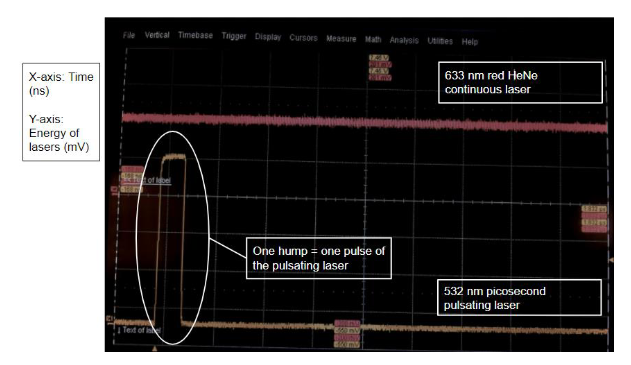}
    \caption{Oscilloscope}
    \label{fig:6}
\end{figure}

A syringe pump and a syringe are attached to one end of the capillary. To start the machine, deionized UV water is added into the syringe and flushed through the microcapillary to clean out any debris from the system. Next, the picosecond laser’s frequency was adjusted to \(50 \, \text{hertz}\). Then, the plasmonic nanoparticle solution is added into the syringe and pumped out the other side with a flow rate of \(6 \, \mu \text{L/minute}\). Finally, the data is recorded and collected on a file via the oscilloscope; the file can then be transported and stored onto a hard drive in a computer.

\section{Plasmonic Nanobubble Counting Mechanism}
The plasmonic nanobubble detection and counting mechanism relies on a couple of key concepts: 
1) The specific wavelength absorbance of \(15 \, \text{nm}\) AuNPs, 
2) the formation of nanobubbles, 
3) light diffraction in nanobubbles, and 
4) the counting mechanism.

The reason why a \(532 \, \text{nm}\) laser was utilized in this experiment was because the optimal light absorbance spectrum for \(15 \, \text{nm}\) AuNPs is around \(520-535 \, \text{nm}\). This means that the AuNPs would have absorbed the greatest amount of energy from lasers within that range, the plasmonic resonance. The laser pulses ensure that the AuNPs will not absorb too much energy and burn up the surrounding microcapillary tube.

Through the properties of surface plasmon resonance, the electrons on the surface of the \(15 \, \text{nm}\) AuNPs will start to oscillate between the poles of the nanoparticle when they absorb the \(532 \, \text{nm}\) wavelength of light from the laser. The electron oscillations will, in turn, cause the AuNP to vibrate and generate lots of heat. The heat will then increase the temperature of the surrounding water, generating a nanobubble around the nanoparticle. The plasmonic nanobubbles referred to in this paper are small vapor bubbles that form when the water surrounding the AuNP gets heated up.

While the picosecond laser is utilized in this experiment as a tool to energize the AuNPs, the \(633 \, \text{nm}\) red HeNe continuous laser is utilized to detect the formation and the sizes of the nanobubbles generated. When a nanobubble is generated, it will diffract some of the light of the red HeNe laser beam (Figure 7), causing the \(mV\) value on the oscilloscope to dip and form a small parabola. 

\begin{figure}[H]
    \centering
    \includegraphics[width=0.5\linewidth]{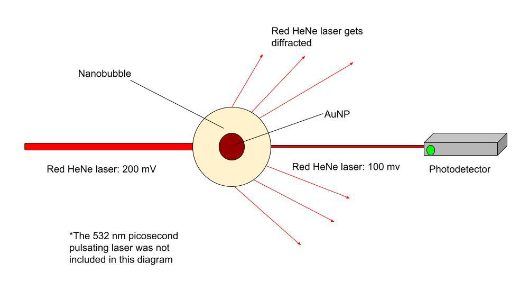}
    \caption{The red HeNe laser getting diffracted by the nanobubble}
    \label{fig:7}
\end{figure}
After every laser pulse, the \(mV\) value of the red HeNe laser beam is collected. If there is no dip of the red HeNe \(mV\) value, then a “Foff” signal is counted; if there is a dip of the red HeNe \(mV\) value, then a “Fon” signal is counted (Figure 8). The number of Fon signals counted is equal to how many AuNPs were detected in the whole solution.

\begin{figure}[H]
    \centering
    \includegraphics[width=0.5\linewidth]{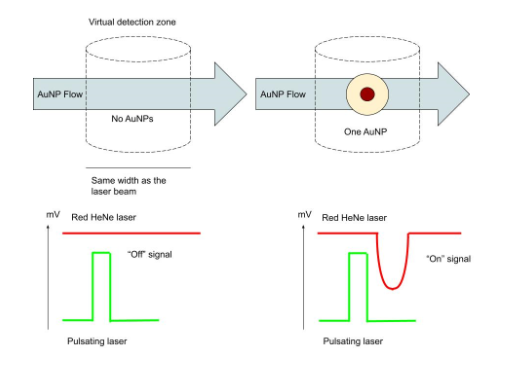}
    \caption{Counting principal for the AuNPs}
    \label{fig:8}
\end{figure}

\section{Single Virion Counting Mechanism}
The basis for the single-virion counting principle builds off of the key concepts discussed in section VII. As seen in Figure 4, after the Synagis antibodies on the AuNP attach onto the F glycoproteins, the AuNPs get aggregated together and form large clumps. When the large clumps pass through the laser’s virtual detection zone, they are all energized by the laser all at once. This, in turn, will form larger bubbles compared to those formed by single AuNPs. The larger bubbles formed by the RSV-conjugated AuNPs will diffract more of the laser beam light compared to the bubbles formed by the single unconjugated AuNPs (Figure 9). 

\begin{figure}[H]
    \centering
    \includegraphics[width=0.5\linewidth]{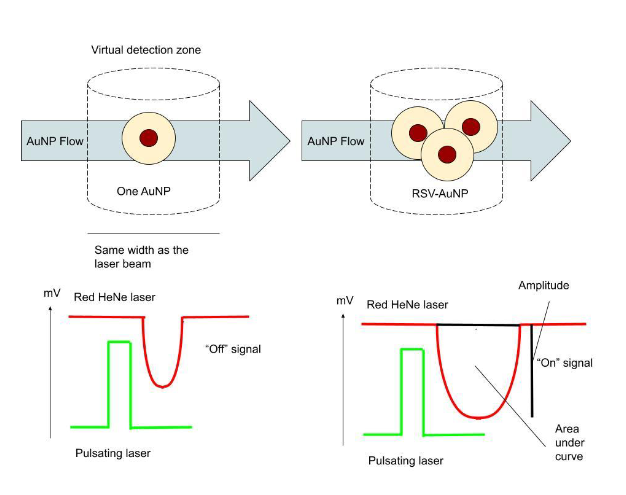}
    \caption{Single AuNP detection vs single virion detection}
    \label{fig:9}
\end{figure}

There is a larger dip for the RSV-conjugated AuNPs compared to the single AuNP detection because as the nanobubbles increase, more of the light energy from the Red HeNe laser will get refracted, indicating a larger dip in the energy (mV) detected by the photodetector.

Additionally, the “Foff” and “Fon” signals were recalibrated and recalculated from the single AuNP detection to fit the RSV-conjugated AuNP detection. To do this, the thresholds were determined by using bivariate data plots to analyze the amplitudes (mV) and the area under the curve (AUC) for the two types of dips: unconjugated AuNP nanobubble dips and RSV-AuNP nanobubble dips. First, the serial dilutions of the unconjugated AuNPs are used as the control group to calculate the thresholds. Next, the thresholds are determined by calculating the mean (\(\mu\)) plus 5 standard deviations (\(\sigma\)), and since both the amplitudes and the AUC of the \(15 \, \text{nm}\) AuNPs are normally distributed, the threshold values covered well over \(99.99\%\) of all amplitudes and AUCs from the control sample. Any amplitude and AUC that is lower than the threshold values will be counted as a “Foff” signal and, any that are higher than the threshold values will be counted as a “Fon” signal (Figure 10). Since the only signals that can possibly have higher AUC and amplitude values than the threshold values are the RSV-AuNP signals, the number of “Fon” signals is equal to the number of virions of RSV counted.

\begin{figure}[H]
    \centering
    \includegraphics[width=0.5\linewidth]{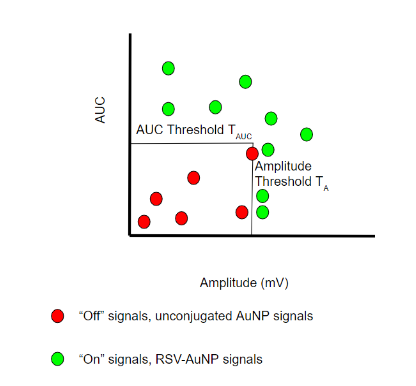}
    \caption{Threshold value determination}
    \label{fig:10}
\end{figure}

\section{Results and Conclusions}
Starting with the single plasmonic nanobubble counting mechanism, the experiment used 6 different \(\lambda\) values (\(\lambda =\) number of estimated AuNPs per virtual detection zone): 0.3, 3, 30, 300 and 0.6, 6, 60, 600. The \(\lambda\) value is determined by \(\lambda = -\ln(1 - \text{Fon\%})\), \(\text{Fon\%} = (\text{Fon}/(\text{Fon} + \text{Foff})) \times 100\). The size of the virtual detection zone is determined by \(V = c/\lambda\), where \(c =\) concentration of NPs. 
\begin{figure}[H]
    \centering
    \includegraphics[width=0.5\linewidth]{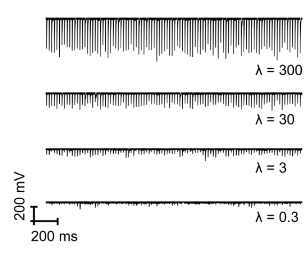}
    \caption{Raw data values for 100 pulses with $\lambda$ between 0.3-300}
    \label{fig:11}
\end{figure}
\begin{figure}[H]
    \centering
    \includegraphics[width=0.5\linewidth]{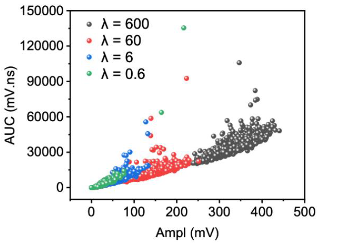}
    \caption{Scatter plot for amplitude and AUC of AuNPs with $\lambda$ between 0.6-600}
    \label{fig:12}
\end{figure}
In Figure 11, each pulse shows a dip in the red HeNe \(mV\) on the oscilloscope. As \(\lambda\) increases, the more AuNPs are in the detection zone meaning that concentration has increased. This has the same effect as the RSV-conjugating AuNP signals as both increasing concentration and conjugation will increase \(\lambda\). As \(\lambda\) increases, the amplitudes and AUC both increased proportionally. In Figure 12, the same experiment was conducted except \(\lambda\) was doubled for every value, and the amplitudes and AUC were extracted from the dips and plotted on a scatterplot.

Next, the single-virion counting mechanism was tested. A \(\lambda\) value of 60 was utilized for this experiment. First, 100 pulses were collected, and their amplitudes and AUCs were plotted on a scatter plot (Figure 13). The thresholds calculated were: \(T_{\text{AMP}} = 225.65 \, \text{mV}\), \(T_{\text{AUC}} = 137.13\). 

\begin{figure}[H]
    \centering
    \includegraphics[width=0.5\linewidth]{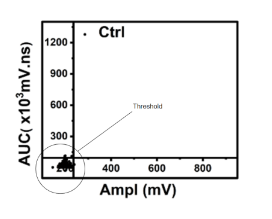}
    \caption{Control sample threshold}
    \label{fig:13}
\end{figure}

\begin{figure}[H]
    \centering
    \includegraphics[width=0.5\linewidth]{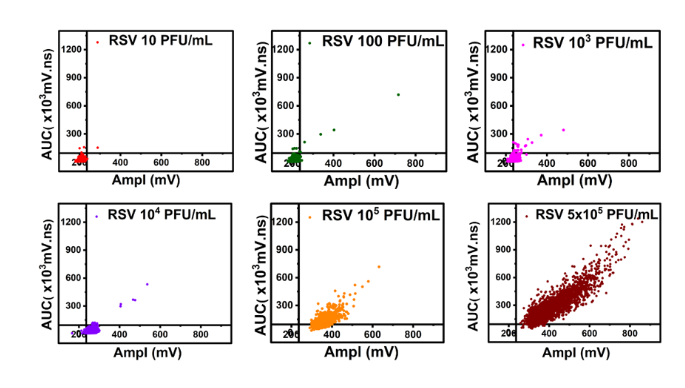}
    \caption{Single virion counting system}
    \label{fig:14}
\end{figure}

As the virion concentration increases, the Fon number increases proportionally too. For example, when the virion sample was at \(10 \, \text{PFU/mL}\), there were only 3 total RSV virions in the entire sample. When the virion sample was increased to \(100 \, \text{PFU/mL}\), the system counted 10 total virions. When the virion sample was increased to \(10^3 \, \text{PFU/mL}\), the Fon\% was 56.45\% with 572 virions, meaning that there were more virions in the sample than free-floating AuNPs. When the virion sample was increased to \(10^5 \, \text{PFU/mL}\), the system determined that the Fon\% of the viral sample was now 100\%, meaning that every AuNP was attached to an RSV virion. The virion counting mechanism can be statistically cross-validated with Poisson statistics.

\begin{figure}[H]
    \centering
    \includegraphics[width=0.5\linewidth]{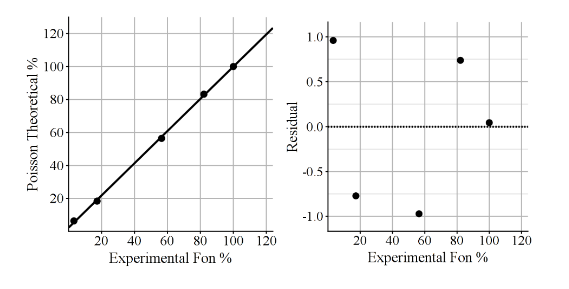}
    \caption{Poisson theoretical \% vs experimental Fon\%}
    \label{fig:15}
\end{figure}

Since there is a randomized residual plot, this means that the relationship between the Poisson theoretical \% and the experimental Fon\% is linear, and the \(r\) value for the scatter plot is 0.989. This shows that the relationship between the two variables is almost perfect. This indicates that there is little to no difference between the theoretical and observed values. The only concern for a slight error could be the \(10 \, \text{PFU/mL}\) titer one as there was a slightly larger difference between the Poisson estimate and the experimental data compared to the other titers.

In conclusion, the machine was very effective in not only diagnosing patients but also counting the number of virions in a fast and rapid way. On average, the test results were retrieved in around \(7\) to \(8\) minutes, faster than our current fastest diagnosis method by half the time while being almost \(30\%\) more accurate. Compared to our current medical standards of LAMP and PCR tests, this test is equally accurate while offering more sensitivity and versatility, all while costing significantly less.

\section{Future Research and Applications}
The ongoing development of this novel system represents a significant leap in public health and urban epidemiology. One of the key challenges we face is managing high viral loads in samples. When the viral load is too dense, our current virion counting system, which relies on AuNPs (Gold Nanoparticles), becomes ineffective due to multiple virions clustering together. To address this, we are exploring the integration of an analog detection system. Although this requires additional time and resources, its potential to enhance accuracy is substantial. A major technical advancement we're considering is the switch from a picosecond to a nanosecond laser for the excitation beam. This change not only promises to shrink the laser's size but also significantly reduce costs, making the technology more accessible, especially in densely populated urban settings where resource allocation is crucial. Furthermore, we are experimenting with varying sizes and concentrations of nanoparticles to optimize the aggregation properties of the AuNPs. This refinement is expected to improve the method's precision and efficiency. In terms of future developments, we working towards the integration of Artificial Intelligence (AI) and Bayesian Ridge Regression into our project. This AI system is designed to revolutionize how scientists approach epidemiology and healthcare policies in urban environments. By leveraging AI, we aim to simplify complex data analysis, making it easier for public health officials to make informed decisions. This approach will enable more accurate predictions of disease spread and effectiveness of intervention strategies, ultimately leading to better health outcomes in urban populations. The combination of these technological advancements and AI integration holds great promise for making epidemiological research and healthcare policy more navigable and effective, especially in the context of the unique challenges presented by urban settings.

\section{Acknowledgement}
This project would not have been possible without the help of all the University of Dallas
research team. We are forever grateful for their resources throughout this entire
project.
\section{Works Cited}
Liu, Y., Ye, H., Huynh, H., Xie, C., Kang, P., Kahn, J. S., \& Qin, Z. (2022). Digital
plasmonic nanobubble detection for rapid and ultrasensitive virus diagnostics. Nature
communications, 13(1), 1687. https://doi.org/10.1038/s41467-022-29025-w

Fajnzylber, J., Regan, J., Coxen, K. et al. SARS-CoV-2 viral load is associated with
increased disease severity and mortality. Nat Commun 11, 5493 (2020).
https://doi.org/10.1038/s41467-020-19057-5

Surface Plasmon Resonance in Gold Nanoparticles: A Review - Iopscience.
https://iopscience.iop.org/article/10.1088/1361-648X/aa60f3.

Li, Wanwan, and Xiaoyuan Chen. “Gold Nanoparticles for Photoacoustic Imaging.”
Nanomedicine (London, England), U.S. National Library of Medicine, Jan. 2015,
https://www.ncbi.nlm.nih.gov/pmc/articles/PMC4337958/.

Shkir, M., Khan, M. T., Ashraf, I. M., Almohammedi, A., Dieguez, E., \& AlFaify, S.
(2019). High-performance visible light photodetectors based on inorganic CZT and
InCZT single crystals. Scientific reports, 9(1), 12436. https://doi.org/10.1038/s41598-
019-48621-3
20

Nguyen, H. H., Park, J., Kang, S., \& Kim, M. (2015). Surface plasmon resonance: a
versatile technique for biosensor applications. Sensors (Basel, Switzerland), 15(5),
10481–10510. https://doi.org/10.3390/s150510481

\end{document}